\documentclass[preprint2,myplots,psfig]{aastex}

\usepackage{graphics}

\slugcomment{ApJL, submitted}

\shorttitle{M31 Microlensing Confirmation}
\shortauthors{Cseresnjes et al.}

\def\ga{\lower 2pt\hbox{$\, \buildrel{\scriptstyle>}\over{\scriptstyle\sim}\,$}}
\def\la{\lower 2pt\hbox{$\, \buildrel{\scriptstyle<}\over{\scriptstyle\sim}\,$}}

\begin{document}

\title{
{\it HST} Imaging of MEGA Microlensing Candidates in M31\footnote{
Based on observations made 1) with the NASA/ESA $HST$, obtained at STScI, which
is operated by AURA, Inc., under NASA contract NAS 5-26555. These observations
are associated with program \#10273,
2) with the INT operated on La Palma by the ING in the Spanish ORM of the IAC,
3) at KPNO, NOAO, which is operated by AURA, Inc., under cooperative agreement
with the NSF.
}}

\author{Patrick Cseresnjes\footnote{Columbia Astrophysics Laboratory, Columbia
University, 550 W.~120th St., New York, NY~~10027},
Arlin P.S.~Crotts,$^2$
Jelte T.A.~de Jong\footnote{Kapteyn Astronomical Institute, University of
Groningen, PO Box 800, 9700 AV, Groningen, The Netherlands},~
Alex Bergier,$^2$
Edward A.\ Baltz\footnote{Kavli Inst.~for Particle Astrophysics \&
Cosmology, Stanford U., PO Box 20450, MS 29, Stanford, CA~~94309},
Geza Gyuk\footnote{Department of Astronomy and Astrophysics, University of
Chicago, 5640 S.\ Ellis Ave., Chicago, IL~~60637},
Konrad Kuijken,\footnote{Sterrewacht Leiden, University of Leiden, PO Box 9513,2300 RA, Leiden, The Netherlands}$^{~,3}$
Lawrence M.~Widrow\footnote{Department of Physics, Queens University, Kingston,
ON~~K7L 3N6, Canada}}

\begin{abstract}
We investigate $HST$/ACS and WFPC2 images at the positions of
five candidate microlensing events from a large survey of variability
in M31 (MEGA).  Three closely match unresolved sources, and two produce only
flux upper limits.  All are confined to regions of the
color-magnitude diagram where stellar variability is unlikely to
be easily confused with microlensing.
Red variable stars cannot explain these events (although background supernova
are possible for two).
If these lenses arise in M31's halo, they are due to
masses $0.15 < m / M_\odot < 0.49$ (95\% certainty, for a $\delta$-function
mass distribution), brown dwarfs for disk lenses, and stellar
masses for bulge lenses.
\end{abstract}


\section{Introduction}

Galaxian dark matter has been recognized 
for over 70 years (Zwicky 1933), and tied in part to the
halo for over 30 (Rubin \& Ford 1970).
Halo dark matter's nature is still a mystery.
Gravitational microlensing can reveal individual
dark matter objects of roughly stellar mass (Paczy\'nski 1986).
To test this, MACHO observed the Magellanic Clouds for 5.7 years, (Alcock et
al.~2000) and EROS (Afonso et al.~2003) did so for 5.
The former report microlensing events more common than the
known, purely stellar expectation, with lensing fraction $f \approx
20$\% of the dark matter halo mass (8-50\%, with 95\% confidence) of
$\sim$0.4 $M_\odot$ masses.
EROS found $f$ consistent with zero (but marginally consistent with $f \approx
20$\%).

M31 microlensing 
could potentially settle this quandary definitively (Crotts 1992).
Since we can explore microlensing across the face of M31, we can use
this distribution to distinguish where in the galaxy the lenses arise.
Several surveys of M31 microlensing 
(Riffeser et al.~2003, Joshi et al.~2004, Calchi-Novati et
al.~2005, including MEGA: de Jong et al.~2004 and its predecessor
VATT-Columbia: Uglesich et al.~2004),
together report $\sim$20 probable microlensing events,
and have a tendency to confirm the MACHO result.

With its crowded target stars, M31 microlensing relies
on image subtraction to reveal event lightcurves, which removes the baseline
flux.
Using $HST$ to recover the source flux (e.g. Ansari et al.\ 1999, 
Auri\`ere et al.\ 2001), one can compute event amplification,
hence Einstein parameters, constraining physical parameters e.g., lens mass.
MEGA and VATT-Columbia also use
source star color to distinguish microlensing from variable stars, since
very red variables (miras and
semiregulars) produce outbursts that, with their baselines
subtracted, mimic point-source, point-lens (``Paczy\'nski'') light curves
(Uglesich et al.~2004).
Residual flux from these events, however, is redder than
almost all potential microlensing source stars.
MEGA will soon publish its microlensing sample, and now is an excellent
opportunity to check these event selection criteria.

\begin{figure*}[]
\includegraphics[width=16.7cm,height=14.75cm]{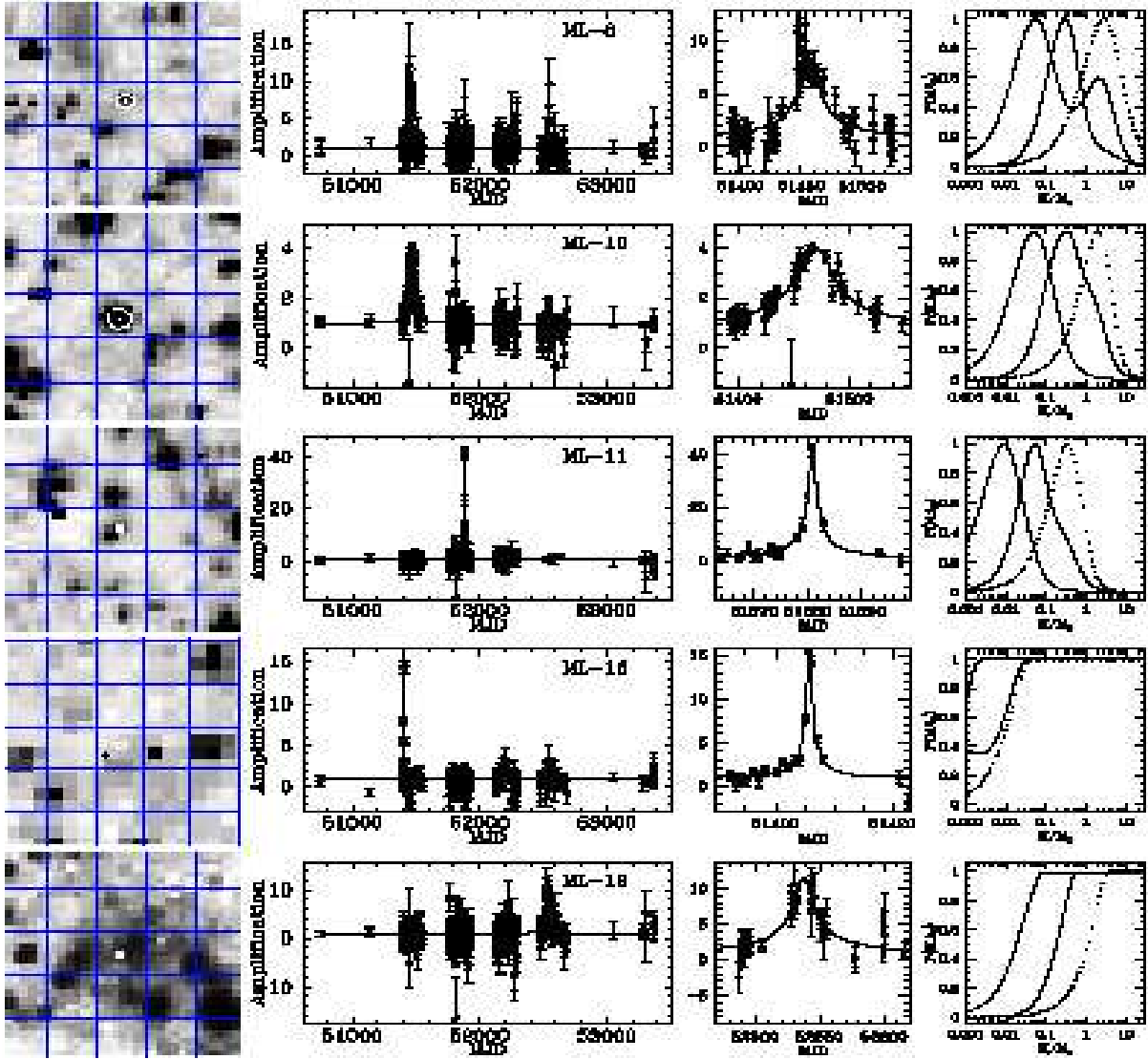}
\caption{Left to right: 1) $(1\arcsec .5)^2$ $HST$ 
image (f814w band) around microlensing candidates. The circles correspond
to $1\sigma$ spread of the individual position estimates. The grid 
represents the INT 
pixel sampling; for ML-16, the cross point corresponds to an independent
estimate as described in the text. 2) full combined light curve 
(Filled squares: KP-R, asterisks: KP-I, open circles: INT-r', 
open triangles: INT-i');
3) Zoom on the event peak; 4) Lens-mass probability distribution 
for a lens in an isothermal halo (full line), in the disk (dashed line), 
and in the bulge (dotted line).}
\label{fig1}
\end{figure*}

\section {Observations and analysis}

\begin{figure}[]
\vskip -0.15in
\includegraphics[width=8cm,height=8cm]{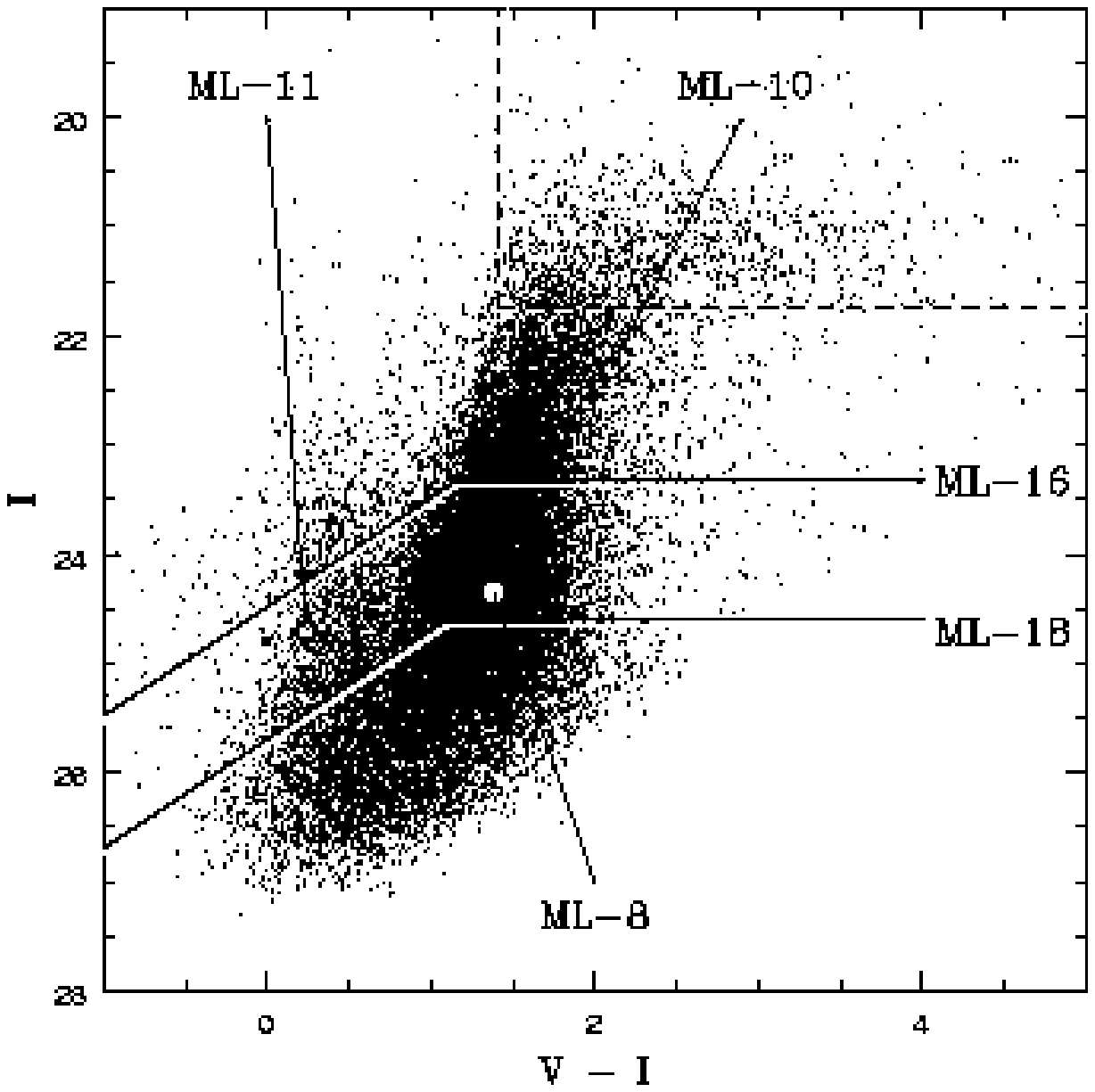}
\vskip -0.3in
\caption{Color-magnitude plot of 5 candidate sources, with
upper flux limits as thick lines. The dashed lines enclose the area where
LPVs and semi-regulars are expected \cite{brown}.}
\label{fig2}
\end{figure}

To study candidate events we appeal to superior $HST$ angular resolution:
160 ACS and WFPC2 images, taken in F555W and F814W filters in 16 orbits,
cover 0.17 deg$^2$, $\sim$30\% of the MEGA field.
Here we study the largest sample of candidate microlensing events, the INT/WFC
subsample of MEGA (de Jong et al.~2005), in order to understand and improve
ground-based selection criteria.

The analysis used is detailed by Cseresnjes et al.\ (in preparation).
We carefully align the $HST$ and ground-based images by matching catalogs of
ground-based versus Gaussian-convolved $HST$ sources
for each filter combination ($HST$, F555W/F814W versus
INT, $r^{\prime}/i^{\prime}$ and KPNO 4m, R/I), providing up to 8
different position estimates.
For a given ground-based position, the two independent $HST$ positions 
(via F555W and F814W) always agree to $\le 0\arcsec .03$ 
(typically $0\arcsec.01$), so
positional accuracy depends mainly on ground-based data.  
The adopted position is a weighted average
of individual estimates (Fig. \ref{fig1}). The spread of different estimates
for each event are $0\arcsec .02$ to $0\arcsec .08$. Of five microlensing
candidates analyzed, we identify three sources and
find flux upper limits for two.

$HST$ data were photometered with DAOPHOT (Stetson 1987), as prescribed in 
Sirianni et al.\ (2005).
The locations of the candidate microlensing sources on a color-magnitude
diagram are shown in Fig. \ref{fig2}.
For each candidate event, we normalized the differential light curves
to R-fluxes, using color-magnitude diagrams and $HST$ baseline fluxes
(for the two undetected events, using the baseline flux upper limit),
then performed a Paczynski fit in ($u_{0},t_{0},t_{E}$)
to the combined light curve (Fig. \ref{fig1}). 
For the two undetected events,
the resulting $t_{_E}$ corresponds to a lower limit.

Only the Einstein time-scale ($t_{_E}$) constrains
lens characteristics, particularly its mass $m$. 
For a given time-scale
$t_{E}$, the lens-mass probability distribution is 
$P(M,t_{E})=({\rm d}\Gamma/{\rm d}t_{E})/\Gamma$, 
where $\Gamma$ is the event rate \cite{griest}.
We consider alternatively a lens located in the halo of M31, 
in the disk or in the bulge. 
For halo lensing, 
we consider the simple case where the lens is part of a
spherical isothermal halo composed of single mass objects, with a density
distribution defined as 
\begin{equation}
\rho_{h} \propto \frac{1}{R^{2}+r_{c}^{2}}
\end{equation}
where $R$ is the radial distance to the center of M31, and $r_{c}$
is a core radius of $5$ kpc. The one-dimensional velocity dispersion
of the lenses is set to $170$ km~s$^{-1}$, consistent with a rotation curve
of $240$ km~s$^{-1}$. The disk is modelled by a double exponential with 
a radial scale length of
$6$ kpc and a vertical scale length of $400$ pc. The bulge corresponds to
the ``small'' bulge model of Kent (1989), with a velocity dispersion
of $150$ km~s$^{-1}$. More details about the model can be found in
Baltz, Gyuk, \& Crotts (2003).

\section{Individual events}

\begin{deluxetable}{lccccc}
\tabletypesize{\scriptsize}
\tablecaption{Event source photometry and Microlensing parameters}
\tablewidth{0pt}
\tablehead{
\colhead{id}         & 
\colhead{ML-8}       & 
\colhead{ML-10}      & 
\colhead{ML-11 (S4)} &
\colhead{ML-16 (N1)} &
\colhead{ML-18}
}
\startdata
R.A. (J2000)         &
00:43:24.53          &
00:43:54.87          &
00:42:29.90          &
00:42:51.22          &
00:43:17.27         \\
Declination (J2000)  &
41:37:50.4           &
41:10:33.3           &
40:53:45.6           &
41:23:55.3           &
41:02:13.7          \\
R$_{hst}$            & 
$24.94\pm 0.14$      & 
$23.36\pm 0.09$      & 
$24.86\pm 0.30$      & 
$>23.86$             & 
$>25.09$            \\
I$_{hst}$            & 
$24.34\pm 0.08$      & 
$22.31\pm 0.07$      & 
$24.71\pm 0.26$      & 
$>$23.32             & 
$>$24.59            \\
(R-I)$|_{hst}$      & 
0.60                & 
1.05                & 
0.15                & 
~~--~~              & 
~~--~~              \\
(R-I)$|_{lc}$       & 
0.59                & 
1.05                & 
0.21                & 
~~--~~              & 
0.51                \\
$A_{\rm max}$       & 
8.49                & 
4.00                & 
41.93               & 
$>$16.01            &  
$>$11.42           \\
$t_{_E}$/day        & 
$60.6\pm 4.2$       & 
$64.7\pm 1.9$       & 
$26.1\pm 1.1$       & 
$>$6.9              & 
$>$86.6            \\
$\chi^{2}/N$        & 
0.89                & 
1.26                & 
1.01                & 
1.29                & 
1.04               \\
$m_{\rm halo}/M_\odot$\tablenotemark{a} &
$0.31^{+0.48}_{-0.21}$ &
$0.33^{+1.04}_{-0.23}$ &
$0.05^{+0.16}_{-0.03}$ & 
$>0.00$                & 
$>0.62$               \\
$m_{\rm disk}/M_\odot$\tablenotemark{a}  &
$0.05^{+1.65}_{-0.03}$ &
$0.05^{+0.09}_{-0.04}$ &
$0.01^{+0.02}_{-0.01}$ & 
$>0.03$                & 
$>0.09$               \\
$m_{\rm bulge}/M_\odot$\tablenotemark{a} &
$2.85^{+4.03}_{-2.31}$  &
$2.00^{+1.66}_{-1.55}$  &
$0.36^{+0.31}_{-0.28}$  & 
$>0.05$                 & 
$>3.73$               \\
comment                & 
red clump or SN        & 
giant branch           & 
very blue              & 
undetected(?)          & 
in cluster or galaxy  \\

\enddata
\tablenotetext{a}{most likely lens mass (with 1$\sigma$ confidence intervals)}
\label{ml_table}
\end{deluxetable}

{\it ML-8:}
this event's position lands within the FWHM of a red clump star's image, 
with $R-I$ in 
excellent agreement with the peak flux's color in
differential light curves ($0.60 \pm 0.16$ vs.\ $0.59$ mag).
With the baseline set
to this star's flux, a Paczynski fit yields amplification $A = 8.49$,
an Einstein time-scale $t_{_E} = 60.6 \pm 4.2$ days. The corresponding
lens masses are $m = 0.31^{+0.48}_{-0.21} M_\odot$ for a halo lens, 
$m = 0.05^{+1.65}_{-0.03} M_\odot$ for a disk lens, and 
$m = 2.85^{+4.03}_{-2.31} M_\odot$ for a bulge lens.\\
\indent However, this event lands $\sim0^{\prime\prime}.9$ from the center 
of a background
galaxy (subtending $\sim1^{\prime\prime}.5 \times 0^{\prime\prime}.3$).
Its color, flux and decline rate are consistent with a Type Ia supernova at
$z \approx 0.5$, with $\la 1$ mag extinction (see Johnson \& Crotts 2005).
One must balance the number of SNe ($\sim 100$ y$^{-1}$ deg$^{-2}$ e.g., Woods
\& Loeb 1998) landing within the FWHM disk of a source star of consistent 
color we would detect
(1.2 arcsec$^{-2}$) versus the number of microlensing events (evidently
$\sim$5) landing so close to an $R<23$ galaxy (100 arcmin$^{-2}$ --- Huang et
al.\ 2001).
The expected number of both kind of events are of the order of a few tenths, 
with perhaps microlensing being slightly more likely.

{\it ML-10:}
this event lands within the FWHM disk of a giant branch
star of color ${\rm R}-{\rm I}=1.05$, in perfect agreement with the
microlensing data.
This source has 
$A = 4.00$ and 
$t_{_E} = 64.7 \pm 1.9$ days, corresponding
to a halo lens mass $m=0.33^{+1.04}_{-0.23} M_\odot$, a disk lens mass
$m=0.05^{+0.09}_{-0.04} M_\odot$, or a bulge lens mass
$m=2.00^{+1.66}_{-1.55} M_\odot$.
It lands suspiciously close to
a region of the CMD common to variables. Still, 
the achromaticity of the variation, the well-fit and
well-sampled peak ($\chi^{2}/N=1.26$), and the stability of the
baseline over $7$ seasons strongly indicate a real microlensing event.
 
{\it ML-11:}
this event lands on a faint blue star 
(${\rm R-I}=0.15\pm 0.40$) severely blended with a red clump star.
The light curves fit yields a similar ${\rm R-I}=0.21$.
Its baseline flux implies $A=41.93$ and $t_{_E}=26.1\pm 1.1$ days. 
This event, from Paulin-Henriksson et al.\ (2002), is
near M32, suggesting that the lens resides there.
If not, the most likely lens mass is $m=0.05^{+0.16}_{-0.03} M_\odot$ in
the halo, $m=0.01^{+0.02}_{-0.01} M_\odot$ in the disk, and
$m=0.36^{+0.31}_{-0.28} M_\odot$ in the bulge.

{\it ML-16:}
this event lands 
in a WFPC2 field and was also seen by POINT-AGAPE (Auri\`ere et al. 2001).
They publish a color for the event peak based on INT g$^\prime$ and 
r$^\prime$ (no i$^\prime$ data are available), corresponding to
V-I~$\sim 2.1$.
We find no detected source at this position; the nearest detected star 
landing $\sim 0\arcsec .1$ away (one WFPC2 pixel) with ${\rm V-I} \simeq 1.1$.
Using the flux of this star as an upper limit, we find $A>16.01$, 
$t_{_E}>6.9$ days,
and $m > 0.003 M_\odot$ being poorly constrained for a halo lens.\\
\indent Auri\`ere et al.\ (2001) seem to have isolated a different source
star. Their celestial coordinates disagree with ours by $3\arcsec$, but
cannot be checked from published data since no image of the source field
is provided. In order to check our astrometry, we repeated the procedure
we applied to r$^\prime$ data to g$^\prime$ data retrieved from the INT 
archive. We also repeated the same procedure (for both r$^\prime$ and
g$^\prime$ data) using the WFPC2 images taken by Auri\`ere et al 
from the archive. Finally, 
one of us (A.C.) made an independent check by choosing $8$ bright 
unsaturated, isolated stars as coordinate inputs to IRAF {\it geomap}
to construct the coordinate transform. These various estimates agree to
better than $0.5$ pixel in the INT data, whereas the nearest star (two
of them actually) with consistent colors and magnitudes to that claimed
by Auri\`ere et al. are $0\arcsec.6$ (or almost two INT pixels) away. Our
position is marginally consistent with the faint source cited above and in 
Table 1, but inconsistent with that of Auri\`ere et al. 

{\it ML-18:}
this event lands in a bright region, perhaps a
cluster or background galaxy.
We isolate no source here, 
so provide only an upper limit baseline flux, estimated by taking the 
brightest pixel within $0^{\arcsec}.05$ and considering that it contains
at most $15\%$ of the source flux, as constrained by the point-spread
function for ACS.
With this flux limit, $A>11.42$, $t_{_E}>86.6$ days, and 
$m\ga 0.62\ M_{\odot}$ for a halo lens, $m\ga 0.09\ M_{\odot}$ for 
a disk lens, and
$m\ga 3.73\ M_{\odot}$ for a bulge lens.

\section {Conclusions}

Of five events in our fields, we find three likely matches, and baseline flux
upper limits on the other two.
Colors of the three identified sources agree with those obtained
from their differential light curves alone.
The two upper limits displace these events from the asymtotic giant branch,
where confusing mira and semiregular variables can occur.
No candidate is a bright red variable.
One might interpret ML-8 as a supernova, but a microlensing
event is just as probable.
We also cannot rule out a supernova as the source for ML-18, which might also
coincide with a background galaxy.
In a future paper, the complete MEGA data set will fill out ML-18's light
curve; unfortunately, we have no additional data on ML-8.

Taking the product of the individual mass probability distributions obtained
for each event, these lenses in 
a halo model (Baltz, Gyuk \& Crotts 2003)
of a single component mass are constrained to $0.15 < m/M_{\odot} <
0.49$ at the 95\% level.
M31 microlensing rates may be consistent with pure
self-lensing (de Jong et al.\ 2005), so we consider bulge lenses 
($0.64 < m/M_{\odot} < 2.02$), or disk lenses which correspond to 
probably unrealistic brown dwarf masses ($0.02 < m/M_{\odot} < 0.06$).

We acknowledge support from STScI (GO 10273) and NSF (grants
0406970 and 0070882).

{}

\end{document}